# Comprehensive Survey: Biometric User Authentication Application, Evaluation, and Discussion

Reem Alrawili, Ali Abdullah S. AlQahtani, *Member, IEEE*, Muhammad Khurram Khan, *Senior Member, IEEE*

*Abstract*—This paper conducts an extensive review of biometric user authentication literature, addressing three primary research questions: (1) commonly used biometric traits and their suitability for specific applications, (2) performance factors such as security, convenience, and robustness, and potential countermeasures against cyberattacks, and (3) factors affecting biometric system accuracy and potential improvements. Our analysis delves into physiological and behavioral traits, exploring their advantages and disadvantages. We discuss factors influencing biometric system effectiveness and highlight areas for enhancement. Our study differs from previous surveys by extensively examining biometric traits, exploring various application domains, and analyzing measures to mitigate cyberattacks. This paper aims to inform researchers and practitioners about the biometric authentication landscape and guide future advancements.

*Index Terms*—Biometric User authentication, Physiological Traits, Behavioral Traits, Cybersecurity in Biometrics, Biometric Traits Analysis, Security Measures in Biometrics

## I. INTRODUCTION

SECURITY is a critical concern in cyberspace applications and services, and user authentication is a crucial step in ensuring the protection of information systems and networks from unauthorized access. As the first line of defense in cyberspace security, user authentication verifies the identity of a user to an authentication entity [1].

Due to the open nature of cyberspace, which exposes the network to a higher risk of security vulnerabilities, user authentication is even more important. Authentication of communication requests originating from unknown or partially known sources has been a popular area of research since the advent of cyberspace communications, with various practical applications [2].

With the increasing use of technology, especially mobile devices, and the exchange of sensitive information [3], user authentication has become more critical than ever. There are currently six factors of authentication that can be used to verify a user's identity: something the user knows, possesses, is, does, the user's location, and something in the user's environment [4]. To ensure the highest level of security, it is crucial to implement a robust user authentication process that incorporates multiple factors [5].

R. Alrawili and A. AlQahtani are North Carolina A&T State University, Greensboro, NC, USA, 27411.
E-mails: rfalrawili@aggies.ncat.edu, AlQahtani.aasa@gmail.com

M.K. Khan is with the Center of Excellence in Information Assurance (CoEIA), King Saud University, Saudi Arabia.
E-mail: mkhurram@ksu.edu.sa

Biometric user authentication has become an increasingly important topic in recent years due to the growing need for secure and convenient identification methods in various application domains. As traditional authentication methods, such as passwords and personal identification numbers, become more susceptible to cyberattacks, biometric systems have emerged as a more reliable alternative. With the rapid advancements in technology, biometric systems have evolved to offer better security, usability, and robustness, making them an integral part of modern-day security solutions. However, as these systems become more prevalent, they also face a range of challenges such as cyberattacks, privacy concerns, and system performance issues.

To address these challenges and to ensure that biometric systems remain effective and reliable, it is essential to have a comprehensive understanding of biometric traits, their application in user authentication, and the factors that impact their effectiveness. Moreover, an in-depth analysis of the advantages and disadvantages of each biometric trait is crucial for the development of improved biometric systems that can cater to various user authentication needs.

This paper aims to provide a thorough review of the literature on biometric user authentication, offering valuable insights into the current state of the field, identifying potential areas for future development, and serving as a valuable resource for researchers and practitioners alike.

In this paper, we answer the following research questions:
1) Which biometric traits are commonly used for user authentication applications, and how suitable are they for specific purposes?
2) How do these biometric traits perform in terms of key performance indicators, i.e., security, convenience, interoperability, scalability, cost, privacy, usability, and robustness?
3) What types of cyberattacks target biometric traits? and how effective are the commonly used biometric traits in upholding the assessment criteria that determine the reliability of a biometric trait?
4) What are the primary factors that impact the accuracy and effectiveness of biometric systems?

The paper's main contributions are summarized as follows:
1) We present a detailed analysis of the biometric traits commonly employed in user authentication. This study assesses their applicability across various settings, establishing a solid foundation for understanding the biometric authentication landscape.

2) Our research extends to a thorough evaluation of these traits, considering key performance indicators, i.e., security, convenience, interoperability, scalability, cost, privacy, usability, and robustness. This approach provides a holistic understanding of each trait's performance in diverse environments.
3) The paper presents types of cyberattacks that specifically target biometric traits. This identification is crucial for understanding the threat landscape and forms the basis for developing effective security measures.
4) We measure the effectiveness of the commonly used biometric traits in upholding the assessment criteria that determine the reliability of a biometric trait.
5) The paper delves into the factors that influence the precision and efficacy of biometric systems. Our discussion sheds light on aspects crucial for the ongoing refinement and advancement of these technologies.
6) We present an objective analysis of the advantages and disadvantages of each biometric trait, identifying areas ripe for improvement and innovation. This balanced viewpoint is instrumental in understanding the current state and future potential of biometric authentication methods.
7) Finally, we provide a visionary perspective on the evolution of biometric authentication techniques. This forward-looking view is essential for guiding future research and development, ensuring that biometric technologies continue to evolve in alignment with user needs and technological advancements.

To answer our research questions, we utilized three main sources: a literature review, online searching, and our own observations. In the first stage, we conducted a comprehensive literature review to obtain detailed information from related existing sources. We searched for relevant articles, and other publications in various databases, including ACM Digital Library, IEEE Xplore, ScienceDirect, Google Scholar, and ResearchGate. We also reviewed the references cited in the selected articles to identify additional sources. To ensure the quality and relevance of the literature we used, we assessed each source based on several criteria, including the author's credentials, the publication venue, the study design, and the date of publication. We extracted relevant information from the selected sources that assisted us in answering all of our research questions and completing our paper.

In addition to the literature review, we conducted online searches to collect more information about biometric authentication techniques. We used various websites, including research databases, industry reports, and google books. We also assessed the quality and relevance of the online sources by considering several factors, such as the website's reputation, the accuracy and reliability of the information, and the source's potential biases or limitations. We used the information we collected from online sources to supplement and validate the findings from our literature review.

In addition to the literature review and online searches, we integrated our personal observations and experiences in the biometric authentication field to enhance our analysis. These observations were derived from our familiarity with diverse biometric systems and our interactions with domain experts, practitioners, and users. We meticulously merged our insights with findings from other sources to provide a comprehensive and balanced perspective, addressing our research questions effectively.

Although we made efforts to collect data from a variety of sources, our methodology has some limitations. For instance, we relied primarily on English-language sources, which may limit the generalizability of our findings to non-English-speaking contexts. We also focused mainly on academic and industry sources. We acknowledge these limitations and encourage future research to address these gaps.

The manuscript is structured as follows: In *Section II*, we provide a comprehensive examination of six essential authentication factors, systematically categorizing biometric traits into physiological and behavioral types, and offer an in-depth analysis of the commonly used biometric traits in user authentication, supported by recent academic research. In *Section III*, we examine the suitability of commonly used biometric traits for specific applications, while *Section IV* evaluates different biometric traits based on various factors. *Section V* discusses biometric traits from different aspects and highlights potential future research directions. Finally, we conclude the paper in *Section VI*. Table I presents all the Tables along with their respective titles.

## II. BACKGROUND

This section explores six pivotal authentication factors essential for verifying user identities. It incorporates a systematic review that organizes biometric traits into two distinct categories: physiological and behavioral characteristics. The section provides in-depth details on the most significant biometric traits employed in user authentication applications for each category, complemented by an analysis of recent academic research related to each trait.

### 1) Something a User Knows

*Something a user knows* is the most commonly used authentication factor in many applications. This refers to the verification process where a user is identified based on information they know, such as passwords, PINs, security questions, and more. However, these pieces of information, commonly utilized as passwords, PINs, or answers to security questions in the realm of cyberspace technology, are easy to remember and weak against potential threats. To mitigate this vulnerability, many organizations have established strict password guidelines, requiring employees to use a combination of upper and lower case letters, numbers, and special symbols. In recent times, instead of relying on strings of characters, passwords can also be graphical, utilizing secret drawings. This alternative approach to authentication is gaining popularity due to its ease of use and potential to increase security.

### 2) Something in a User's Possession

The combination of an individual's possession of a known hardware device and their knowledge creates a secure means of identifying a user, also known as Object-Based Authentication. This authentication factor has gained popularity due to the portability of devices like smartphones, USB keys, tokens, and smart cards. This factor provides an extra layer of security by combining the hardware device with something the individual knows, such as a PIN or password. Even if the password or other information is obtained by a third party, unauthorized access to sensitive information can be prevented. Furthermore, hardware devices can now incorporate biometric authentication such as fingerprint or facial recognition, adding another layer of security to the authentication process with technological advancements.

TABLE I: List of Tables

| Table | Caption |
|---|---|
| Table II | Suitability of Biometric Traits for Different Authentication Applications |
| Table III | Comprehensive Evaluation of Authentication Traits Based on Assessment Criteria |
| Table IV | FAR and FRR Comparison Across Biometric Traits |
| Table V | Comparative Analysis of Biometric Traits Across Key Performance Indicators |
| Table VI | Overview of Cybersecurity Threats to Biometric Traits |
| Table VII | Comparative Impact Analysis of Influential Factors on Different Biometric Traits |
| Table VIII | Summary of Advantages and Disadvantages: Physiological Traits |
| Table IX | Summary of Advantages and Disadvantages: Behavioral Traits |
| Table X | Comparative Coverage Analysis of Biometric Traits in Our Survey and Recent Published Surveys |
| Table XI | Comparative Overview of Evaluation Concepts in Our Survey and Recent Published Surveys |

*3) Something a User Is*

*Something a user is* refers to the unique biometric traits of a user, such as fingerprints, iris patterns, and facial features, that serve as proof of identity. This authentication factor is highly secure since it cannot be transferred or physically stolen, unlike passwords and security tokens. Biometrics-based authentication has gained popularity due to its reliability and convenience, and it plays a crucial role in ensuring the privacy and security of both individuals and organizations. The biometric data used in this method is unique to each user, making it nearly impossible to falsify or duplicate. As a result, it is widely recognized as a key component in secure identity verification processes.

*4) Something a User Does*

Individuals exhibit unique behaviors that can be observed when performing a small task, such as typing or moving. These unique behavior patterns can be captured and utilized to identify individuals through a method known as *Something a User Does* authentication factor. This authentication factor recognizes the distinctive patterns and regularities in an individual's behavior, such as typing rhythm and mouse movement, for authentication purposes. The utilization of *Something a User Does* authentication factor in identifying individuals is crucial for ensuring the security and privacy of sensitive information in various domains, including finance, health, and government. It is increasingly being used as a secondary or multi-factor authentication method to enhance the security of traditional password-based authentication systems.

*5) Somewhere That a User Is*

Location-based authentication, also known as *Somewhere that a user is* authentication factor, is a method of verifying an individual's identity by using their location. This authentication factor has been studied and reported in various research works such as [6] and its idea is to detect the user's location through various mechanisms and then use that information to verify their identity. One common method of determining a user's location is through the use of the Global Positioning System (GPS) and Wireless Fidelity (Wi-Fi), which triangulate the user's position. Geolocation is another mechanism for establishing a user's location that relies on datasets of Internet Protocol (IP) addresses. By combining these various location detection mechanisms, a highly accurate and unique location for a user can be established and used for authentication purposes.

*6) Something That is in The User's Environment*

Recently, researchers have explored using unique environmental factors to authenticate a user's identity, fueled by the prevalence of smart devices that can detect radio frequency signals such as beacon frames and RSSI values [7]. The advantage of these signals is that they can be processed by devices without needing a connection to the broadcaster, while their limited range enables proximity-based applications [8]–[11]. Ambient sounds can also be captured by smart devices and compared across devices for similarity, such as cellphones and laptops [12]. Consequently, authentication based on unique environmental factors, like ambient sound and access points, has emerged [13], [14].

Biometrics authentication is a highly reliable mechanism that offers significant advantages over other authentication factors in many applications. Studies such as [15]–[20] have shown that biometrics play an essential role in increasing user privacy. One significant advantage of biometric technologies is that they do not require users to memorize or carry items during the authentication process. This feature makes biometric access highly reliable, as biometric characteristics are always with the user and cannot be forgotten or stolen.

There are two categories of biometric characteristics: physiological and behavioral. Physiological characteristics are related to the user's physical traits, while behavioral characteristics are related to the user's behavior and habits.

*A. Physiological Traits*

The physiological traits of an individual's body, including their form, shape, and size, are closely linked to user authentication [21]. Among these traits, fingerprint recognition, face recognition, iris recognition, heartbeat recognition, vein recognition, ear recognition, hand geometry, and retina recognition are the most widely used. In the subsequent sections, we will provide a thorough examination of each of these physiological traits.

*1) Fingerprint Recognition:* Fingerprints are a widely used method for identifying individuals due to their unique nature. The National Forensic Science Technology Center (NFSTC) confirms that even identical twins will have different fingerprints [22]. The use of friction ridge skin impressions or fingerprints for identification dates back to 300 B.C. in China and A.D. 702 in Japan, and has been used in the United States since 1902 [23]. In addition to traditional identification methods, fingerprints have also been employed in cyberspace for authentication purposes. Any device with a touchable surface can serve as a convenient fingerprint scanner, providing a suitable environment for fingerprint authentication mechanisms.

Fingerprint recognition technology is extensively researched and implemented in biometric authentication. The development of a dual authentication system using fingerprints and passwords improved security and accessibility in laboratory rooms [24]. Enhancing user experience, a method to reduce response time in fingerprint systems was proposed with 95.83% accuracy in a FoD image dataset [25]. Deep learning algorithms enabled recognition of fingerprints and signatures [26], while ATM machines benefited from secure and convenient transactions [27]. Amidst COVID-19, students authenticated for exams using fingerprint sensors and temperature recording [28]. A secure DHCPv6 system integrated MAC address whitelist authentication and terminal fingerprint identification for campus networks [29].

Security concerns were addressed in [30], and a secure IoT authentication system was proposed with a two-stage feature transformation scheme [31]. The Secure Password-driven Fingerprint Biometrics Authentication (SPFBA) method [32] offered high accuracy for smartphone security compared to other methods. A novel technique that considered minutiae points, ridge count, and local minutiae point structure enhanced fingerprint-based biometrics protection [33]. Overcoming traditional weaknesses, a multi-features fusion and deep learning-based algorithm [34] was proposed, along with a smartphone-based system featuring Wi-Fi communication and cryptography techniques, achieving a 96% recognition rate [35].

*2) Face Recognition:* Face recognition technology has come a long way since its inception in the mid-21st century. Initially, it was limited to identifying traits associated with eyes, ears, nose, mouth, jawline, and cheek structure. However, with the advancement of technology, facial recognition has now evolved to rely on sophisticated mathematical models and comparison processes that use both random (feature-based) and photometric (view-based) features. The technology works by analyzing the structure, shape, and proportions of a person's face, including the distance between their eyes, nose, mouth, and jaw. It also examines the upper outlines of their eye sockets, the sides of their mouth, the location of their nose and eyes, and the surrounding area of their cheekbones. These features are matched with datasets to provide accurate identification [36].

Facial recognition in biometric authentication has been widely researched for its security and convenience. Various methods have been proposed to enhance accuracy and security, such as secret-sharing-based models for public transportation systems [37], incorporating users' psychophysiological states [38], and two-factor systems using QR codes [39]. EchoPrint [40] and electronic voting systems [41] are other secure applications, while some studies suggest augmenting traditional authentication methods with facial biometrics [42]–[44].

Deep learning has been employed for self-adapted verification [45] and flexible examinee verification [46]. Face identification techniques have numerous applications in computer vision, including surveillance [47]–[49], with deep learning approaches like CNN and Softmax classifier improving accuracy [50], [51]. Thermal imaging has been used in personal identification and emotion analysis [52], [53].

*3) Iris Recognition:* Iris recognition is a highly reliable biometric authentication technique that identifies individuals based on the unique patterns found in the annular region surrounding the pupil. This area contains intricate details such as crypts, radial grooves, threads, pigment frills, spots, stripes, and arcuate ligaments that result in a distinctive and unique pattern for each individual. The asymmetrical and randomly dispersed form of the iris makes it one of the most dependable biometric traits, with each eye having its own unique pattern that is not duplicated in the same individual's eyes. As a result, iris recognition is a valuable and widely applicable tool in the field of user authentication, as confirmed by various studies such as [54].

Iris recognition, a widely used biometric authentication method, was first proposed by John Daugman [55] due to its non-contact nature and low likelihood of change. However, conventional systems have time efficiency limitations [56]. To address this, a dynamic system using high-speed image processing was introduced [56], while concerns about feasibility and variability reduction led to the development of a microprocessor-FPGA platform combined with a hardware-friendly error correction code scheme [57]. Enhanced security during the COVID-19 pandemic inspired a dual-factor authentication system combining iris and electrocardiogram (ECG) features [58].

Moreover, bio-cryptosystems using biometrics and encryption in information security are gaining traction [59]. A multi-state iris multi-classification recognition method based on a convolutional neural network fusion statistical cognitive learning was proposed for various acquisition states [60]. The suitability of Head-Mounted Displays (HMD) as a platform for iris recognition was also evaluated [61]. New iris authentication methods, including gray-level co-occurrence matrices and Haralick features, have been proposed [62]. Moreover, iris recognition and One-Time Passwords (OTP) were suggested as additional security measures in a blockchain-based land record system [63]. Lastly, a high-efficiency iris segmentation approach, IrisParseNet, was proposed to tackle noise in non-cooperative iris recognition, showing state-of-the-art results [64].

*4) Heartbeat Recognition:* Heartbeat recognition is a promising biometric trait that relies on capturing cardiac gestures through non-invasive measurement techniques such as Electrocardiogram (ECG), Photoplethysmogram (PPG), Seismocardiogram (SCG), and Phonocardiogram (PCG). These signals are acquired from the surface of the individual's body and reflect the intricate interplay of sympathetic and parasympathetic factors within the human body [65]. Compared to traditional biometric traits like fingerprints and faces, the heart waveform is more complex and challenging to control or alter, which makes heartbeat recognition an effective tool for user authentication mechanisms.

Heartbeat recognition in biometric authentication has gained significant attention, with electrocardiogram (ECG) signals emerging as a reliable modality. Addressing the duration limitation, various methods have been proposed for single heartbeat recognition. For example, [66] used a pre-trained convolutional neural network for ECG signal classification, achieving up to 99.94% accuracy. [67] introduced a hybrid deep learning model considering both local and long-distance heartbeat information, with accuracies up to 99.32% and 97.15%. [68] employed wave modeling for feature extraction, while [69] used a point process model for emotion characterization, achieving 77% accuracy.

[70] explored ML methods for telemedicine applications, and HeartPrint [71] demonstrated an average authentication accuracy over 95% using skin surface vibrations. [72] introduced double authentication with photoplethysmogram and ECG signals for extra security, achieving 99.98% accuracy. [73] proposed a deep learning-based identification scheme without R-peak detection, reaching a 99.1% identification rate. Lastly, [74] compared three ECG-based authentication methods, with deep learning yielding the lowest false acceptance and rejection rates.

*5) Veins Recognition:* Vein patterns located beneath the skin are a unique biometric trait that cannot be easily altered or duplicated. These patterns can only be detected and observed using Near Infrared (NIR) sensors. This is because the veins in an individual's body spread in distinct patterns and configurations. Vein recognition, which utilizes this biometric method, is commonly used for verification purposes. It has been proven to provide a consistent means of user authentication, as supported by several studies [75], [76]. Given the uniqueness of vein patterns from one person to another, this biometric method is highly reliable and secure.

Biometric authentication systems have gained popularity in recent years, with finger vein patterns receiving significant research attention due to their non-intrusive and reliable nature. Various studies have proposed personal authorization methods using these patterns [77]–[88]. Methods include near-infrared (NIR) palm vein pattern recognition using image processing and machine learning [77], portable finger vein recognition with System on Chip (SOC) solutions [78], enhanced finger vein algorithms based on band limited phase only correlation (BLPOC) [79], improved weighted sparse representation for low-quality images [80], and multi-mode local coding operator fusion for increased feature extraction [81].

Other approaches include dorsal hand vein recognition with statistical and Gray Level Co-occurrence Matrix (GLCM) features and artificial neural networks (ANN) [82], end-to-end Convolutional Neural Network (CNN) models for palm vein recognition [83], finger vein recognition with Quadrature Discriminant analysis and Minimum Distance Classifier [84], and a novel acquisition mechanism based on vein pulsation [85]. Cross-sensor finger vein recognition has also been studied, evaluating the performance with different devices [86], wrist hand vein image recognition using the Local Line Binary Pattern (LLBP) method [87], and an overview of various biometric identifiers, including finger vein identification as a new methodology [88].

*6) Ear Recognition:* Decades of research in anthropometrics have confirmed that ear photographs possess unique characteristics in terms of their form and shape, even for identical twins. This phenomenon is referred to as ear recognition. Law enforcement agencies have adopted ear recognition as an essential tool in forensic science for identifying individuals. The recognition process involves capturing either two-dimensional or three-dimensional images of the ear, as the pinna and ear structure are distinct for every individual. The shape of a person's ear remains constant and does not undergo significant changes with age, thus making ear recognition a promising authentication mechanism [89].

Ear recognition has gained popularity in biometric authentication due to its unique and stable features. The pinna's acoustic transfer function (PRTF) is studied for authentication, but its accuracy decreases with positional fluctuation [90]. This issue is addressed by combining PRTF, images, and positional sensor information for robust multimodal authentication. A reduced dimension feature vector concatenation method is proposed for ear, face, and palmprint biometrics, using one algorithm for feature extraction and Euclidean distance for matching to lower computational complexity [91]. Ear and arm gesture as complementary biometrics have been tested on a database of over 100 subjects [92].

Various techniques for feature extraction and classification are employed, such as histograms of oriented gradients and local binary patterns [93], 2D ear imaging with the YOLO algorithm [94], and a combination of DCT, PCA, and SVM [95]. PRTF is combined with acceleration and proximity sensor information for improved robustness [96], and a real-time ear detection method using CenterNet deep learning network is proposed [97]. Ear-based authentication systems utilize CNN, morphological postprocessing, and discrete Radon transform for feature extraction [98].

Ear recognition, combined with face biometrics, achieves a 97.1% recognition rate for border control [99]. Also, studies apply both texture and geometric features [100], a cascaded classifier-based ear detection approach with Shape Context descriptor [101], and a Match Region Localization Ear Recog-nition System [102], all demonstrating promising results for ear recognition in biometric authentication.

*7) Hand Geometry:* The human hand is a valuable source of unique information, with its distinct characteristics such as shape, length, width, thickness, and palm line pattern that can be used for identification purposes through hand geometry recognition. These features are unique to each individual and cannot be easily discerned by the naked eye. However, electronic instruments such as cameras can capture and identify them. The process of hand geometry recognition involves scanning the hand's size and shape, as well as capturing its length, width, thickness, and surface area through a camera [103]. This information can be utilized to authenticate an individual and ensure a secure exchange of sensitive information.

Hand geometry has recently gained traction as a biometric modality for authentication, with various aspects such as palm geometry, palmprint, and hand structure being investigated [103]–[105]. Some studies propose multimodal biometric systems combining iris and hand geometry features ( [106]) or utilizing dorsal and palmar hand images [107]. Others explore unique approaches like acoustic signal scanning [108] or hand radiograph-based authentication [109]. Continuous authentication systems like HandPass [110] and gesture recognition techniques using CNN-RNN methods [111] have also been developed. CNN-based hand gesture detection [112] and user-independent hand gesture recognition systems [113] have been proposed, with the latter achieving superior performance on the American Sign Language (ASL) benchmark dataset.

Research has also focused on hand image-based gender recognition and biometric identification, introducing the 11K Hands dataset for various tasks [114]. GPA-Net, which learns global and local deep feature representations, further improves hand-based person identification in large, multi-ethnic datasets [115].

*8) Retina Recognition:* The human retina is a delicate tissue located on the posterior surface of the eye. It exhibits a unique structure and pattern of blood vessels for every individual unless affected by any pathological condition. This exceptional arrangement is referred to as retina recognition, which can be evaluated using retinal scanning technology. Retinal scanning captures digital images of the retina, highlighting its individuality [116]. The singularity of the retina has made it an important tool for biometric authentication. Biometric authentication research has extensively explored retina recognition, focusing on segmentation as a critical step for reducing false recognition rates.

A multi-modal system fusing iris and retina recognition, along with PCA for addressing the curse of dimensionality, achieved a 98.37% recognition rate [116]. Improved algorithms using the AKAZE detector, FREAK descriptor, and FPGA implementation increased repeatability to 90.9% [117]. The LHPB binary descriptor introduced in [118] improved accuracy by 17% compared to Chen et al.'s method. Retina-based pseudorandom number generators have been developed for security applications [119], while the highest diagnosis accuracy (96.45%) was obtained using a third-degree polynomial SVM for retinal characteristic point extraction [120]. However, inverse biometrics can compromise personal information in biometric systems [121]. Eye movement dynamics, though promising, face challenges in long-term experiments and require further consideration of visual tasks and eye trackers' resolution [122].

Liveness detection in retina recognition has been addressed using laser speckle contrast imaging [123]. Quantum vision introduces photon counting in the human visual system for biometric recognition [124]. Radial Chebyshev Moments (RCMs) shape descriptors, PCA, and an SVM classifier have been employed for blood vessel segmentation, achieving high identification rates and varying processing times [125]. Retinal recognition has been compared to fingerprints and iris patterns, while the impact of diseases on retinal recognition has been investigated [126]. Finally, a supervised method combining CNN and RF for retinal blood vessel segmentation showed promising results on DRIVE and STARE databases [127].

*B. Behavioral Traits*

Behavioral biometrics emerge as an alternative way of identification. User authentication via behavioral biometrics is relatively less established than the use of physiological biometrics. This concept involves a range of gestures including those normally applied to mobile devices such as touch gestures, voices, motion and orientation as well as others such as mouse dynamics, handwriting, and gait [128].

In contrast to biometric traits that rely on physical attributes to verify an individual's identity, behavioral traits confirm an individual's identity through unique patterns displayed during interaction with a device. These patterns include signature recognition, gait recognition, voice recognition, and keystroke and mouse movement, and can be used as methods for user authentication. In this section, we provide examination in details of the major behavioral traits.

*1) Signature Recognition:* Signature recognition is a well-known physical activity that is utilized for identification purposes in several sectors, such as health and finance. There are two forms of signature recognition used for user authentication: static and dynamic. Static signature recognition involves converting an individual's handwritten signature into an image, which is then compared with previously stored signatures. On the other hand, dynamic signature recognition is a more advanced mechanism that captures various dynamic features of an individual's signature, including direction, stroke, pressure, time, and shape, when signing on a touchable surface of smart devices [129].

Signature recognition serves as a prevalent behavioral biometric for personal authentication, but current systems often need five or more signature samples for accuracy [130]. Research has explored various approaches, such as an SNN-based method achieving 84% accuracy [130], a convolutional neural network for off-line recognition with few training samples [130], and a single signature duplication scheme (SRSS) with comparable performance to traditional methods [131]. Cryptographic algorithms like GOST and RSA have been suggested for enhanced security [132]. A CNN-based

7technology using four datasets and N individuals with M signatures demonstrated potential [133].

A novel Siamese network focusing on stroke features reached 82% accuracy on the SigComp2011 dataset [134]. Exploring individual sampling frequencies led to a verification system with accuracy improvements of 70% and 92% in different scenarios [135]. Deep Generative Adversarial Networks (DGANs) showed promise in learning online signatures and higher classification accuracy in discriminating real and fake samples [136]. A modified GOST digital signature algorithm reduced computation complexity by 50% for faster verification [137].

*2) Gait Recognition:* Gait recognition, a biometric authentication method, identifies individuals by their walking style that is shaped by the body's anatomy. Each person has a distinct gait pattern that is difficult to replicate. Gait recognition has emerged as a biometric feature that can be used to identify people in video surveillance systems.

Various studies have explored ways to enhance recognition rates. One of the key advantages of gait recognition is that it does not require active participation from the person being identified, making it a promising area of research in the field of intelligent systems [138]. [139] developed a deep Convolutional Neural Network (CNN) architecture, demonstrating competitive performance compared to previous subspace learning methods. [140] proposed a feature selection mask, yielding a 77.38% correct recognition rate.

PoseGait, a novel model-based method using 3D pose estimation, exhibited state-of-the-art performance and robustness to view and clothing variations [141]. [142] introduced a gait recognition method resilient against intra-subject posture changes using a deformable registration model. [143] presented a deep auto-encoder-based algorithm for gait recognition from incomplete cycle information. [144] suggested a technique dividing gait cycles into phases and converging feature information using Constrained Fuzzy C-Means. [145] proposed multimodal fusion-based gait feature identification algorithms, displaying improved recognition accuracy and robustness.

[146] developed a Multi-view Gait Generative Adversarial Network (MvGGAN) to extend gait datasets for deep learning-based cross-view methods, improving performance. [147] proposed a gait phase recognition algorithm using sEMG of human lower limbs and an improved KNN-dagsvm fusion algorithm, achieving a 95.00% average recognition rate. [148] demonstrated a high-performing regional-LSTM learning model with a 98.47% average recognition rate. Lastly, [149] introduced the Gait Optical Flow History Image (GFHI) method, providing a comprehensive motion representation, and a combined Hidden Markov Model (HMM) for gait recognition, resulting in high accuracy across single, dual, and three-view training data.

*3) Voice Recognition:* Voice biometrics have gained recognition as a top tool for identity verification, with many administrations utilizing it as an authentication factor. The uniqueness of each person's voice, with differences in pitch serving as a reliable identifier for individuals, is the primary reason for this trend [150]. Moreover, voice recognition has proved to be a valuable asset in criminal investigations, where a voice sample can be used to identify suspects even when other identifying factors are unavailable.

Voice recognition is a crucial aspect of biometric authentication, offering convenience and accessibility without requiring special devices [151]. Researchers focus on improving the quality of voice authentication systems, analyzing the informativeness and stability of user templates based on phase data [152] and addressing long-term speaker variability [153]. To combat telecommunications fraud, a dual identity authentication mechanism has been proposed, combining the Hermes algorithm and voice-print recognition technology [154].

Multimodal biometric authentication, which includes face and voice recognition, provides better security [155]. Voice biometrics, however, are susceptible to replay attacks [156]. Solutions like VoiceGesture and automatic emotion recognition systems have been proposed to enhance security [156], [157]. Other approaches include the Gaussian Mixture Model (GMM) for voice authentication [158], the VOGUE system for wearable devices [159], and a novel software-only anti-spoofing system for smartphones [160]. To further protect voice assistants, methods using Microsoft Azure Speaker Recognition API and Google Speech API have been proposed to ensure resistance to attacks and ease of use [161].

*4) Keystroke and Mouse Movement:* Keystroke dynamics and mouse movement analysis are biometric authentication methods that rely on a person's unique typing or mouse movement patterns to identify them. These patterns are analyzed based on various aspects such as typing speed, rhythm, time interval between key presses, speed, acceleration, and direction. To analyze typing patterns, factors such as fly time, press time, and dwell time are taken into consideration. Fly time refers to the time it takes for a user to move their finger from one key to another, press time refers to the time it takes for a user to press and release a key, and dwell time refers to the time a user's finger remains on a key after pressing it [162]. These factors, along with others, are used to create a unique profile for each user, which can then be utilized for authentication purposes.

Keystroke and mouse movement biometrics are crucial in user authentication research. Time series-based methods improve performance in real-time authentication using keystroke dynamics [163]. Adapting models to individuals enhances emotional state recognition and user authentication [164]. Novel algorithms and technologies, such as smart rings and deep learning techniques, enable faster and more accurate keystroke recognition, especially for mobile devices [165]–[167]. Clustering-based anomaly detection techniques achieve the highest AUC in user authentication using keystroke dynamics [168]. A three-step authentication model incorporating mobile orientation and accelerometer data has been proposed for mobile phones [169].

Mouse dynamics-based biometric authentication faces challenges with publicly shared datasets like the Balabit dataset due to short test sessions [170]. A comprehensive review of behavioral biometric strategies, including keystroke analysis and mouse dynamics, is provided by [171], while [172] presents a mouse-based authentication scheme called MAUSPAD. Remote keylogging attacks on search engine autocomplete and fusion keystroke time-textual attention networks for continuous authentication have been explored [173], [174]. Finally, a piezoelectric touch sensing-supported keystroke dynamics-based technique offers secure smartphone access [175].

## III. APPLICATION SUITABILITY

This section discusses the suitability of commonly used biometric traits in various user authentication applications.

### A. Static Authentication Application

Static authentication applications are those that require authentication at a single point in time, without the need for constant monitoring [176]. Among the discussed biometrics in the table, fingerprint, face, iris, veins, ear, hand geometry, retina, and signature are suitable for static authentication applications. However, fingerprint, face, and iris recognition are the most commonly used biometric modalities for static authentication applications, as they are easy to capture and have high accuracy rates.

Moreover, heartbeats, gaits, voiceprints, and Keystroke and mouse dynamics are not suitable for static authentication applications, as they require continuous monitoring and are subject to change. Empirical evidence suggests that static authentication is effective in identity verification. For instance, a study conducted by [177] demonstrated a high level of accuracy in fingerprint recognition, with a false acceptance rate (FAR) as low as 0.01%.

### B. Remote Authentication Application

Remote authentication application is a method of verifying the identity of a user accessing a network or system remotely, such as through the internet [178]. Among the discussed biometrics traits, only heartbeat, signature, and voice, keystroke and mouse movement are suitable for remote authentication applications. The heartbeat biometric can be measured remotely using a wearable sensor, making it an ideal option for remote authentication. Signature biometrics can also be captured remotely using a stylus or touchpad, and voice biometrics can be analyzed from a phone call or a standard microphone. Finllay, keystroke and mouse movement are also suitable as they can be captured using and a keyboard and a mouse.

On the other hand, biometrics such as fingerprint, face, iris, veins, ear, hand geometry, and retina are not suitable for remote authentication applications since they require physical proximity to the scanning device for accurate measurement. Several remote applications have been designed to recognize and prevent crimes, including acts of violence, have been developed through live recordings of security cameras [179], supporting authorities to address instances of risk and threat to security [180].

### C. Dynamic Authentication Application

Dynamic authentication applications rely on continuous monitoring or updates to the biometric data. These biometric measurements are then compared to the user's baseline biometric data, which was established during the initial authentication process [181]. If the measurements match the baseline data, the user is authenticated. However, if the measurements do not match the baseline data, the user may be prompted to re-authenticate or may be denied access.

Among the discussed biometrics traits, only gait analysis, heartbeat analysis, and keystroke and mouse are suitable for dynamic authentication applications. For example, gait analysis can be used to authenticate users based on the way they walk. The system continuously monitors the user's gait and compares it to the previously recorded data to determine the user's identity. Another example of a dynamics user authentication method was combining keystroke dynamics data and image data with artificial image data, using AlexNet and ResNet to classify the artificial image, and achieving an accuracy of 98.57% [182]. Similarly, heartbeat analysis can be used to authenticate users based on their unique heart rate patterns.

### D. Real-time Authentication Application

Real-time authentication applications utilize biometric traits to verify the identity of a user immediately. Biometric trait includes fingerprints, face recognition, ear recognition, hand geometry, retina recognition, signature recognition, and voice recognition. These biometrics are suitable for real-time authentication applications since they provide quick identification of users [183]. For example, in a real-time authentication system that uses fingerprint recognition, the system captures the fingerprint of the user and compares it to the stored fingerprint data in the system's database. If there is a match, the user is authenticated, and if not, access is denied. In a real-time application, face recognition was used and achieved 99.32% as an accuracy [184].

### E. Continuous Authentication Application

Continuous authentication application that uses biometrics traits in a user authentication system works by continuously verifying a person's identity while they are interacting with a device or system. In a continuous authentication system, a user's biometric data is captured and analyzed in real-time to ensure that the person interacting with the device is the same person who was authenticated initially [185].

While biometric traits exist, not all are suitable for continuous authentication. For instance, fingerprint and facial recognition technologies are not suitable for continuous authentication since they only authenticate the user at the point of login. However, biometric traits such as gait analysis, heartbeat analysis and keystroke analysis are suitable for continuous authentication since they can be continuously analyzed. In continuous authentication, head motions was used to classify an authorized and unauthorized user. The system performs better by attaining the accuracy of 99 % in 0.02 seconds providing a continuous authentication to the user [186].

### F. Contact-Based Authentication Application

Contact-based authentication applications utilize biometrics to authenticate users through physical contact with the authentication device. One example is fingerprint authentication, where the user's fingerprint is scanned and compared to a stored template for authentication [187]. Hand geometry authentication is another example, where the user's hand is placed on a scanner to measure the size and shape of their hand for authentication. Hand images have been used in contact-based authentication and achieved 100% recognition rate with normal conditional images, where, different conditional images attained 97.88% recognition rate [188].

### G. Contactless-Based Authentication Application

Contactless-based authentication applications are those that do not require physical contact between the user and the authentication device [189]. One example of a contactless biometric authentication method is facial recognition, which uses the user's face to verify their identity. The system captures an image of the user's face and analyzes it against the pre-existing image stored in the database. If the system identifies a match, the user is granted access. The palm vein pattern biometric have been utilized to authenticate users in a contactless-based authentication and achieved an accuracy of 99.53% [190].

Table II presents the the commonly used biometric traits in user authentication applications and their suitable applications. From this table, it is clear that different biometric traits are used for different applications and the type of biometric trait used depends on the type of application. This highlights the importance of choosing the right biometric trait for the right application to ensure its effectiveness and accuracy. The use of biometric authentication mechanisms is expected to increase in the future, as the technology continues to advance and more applications become available [191].

## IV. EVALUATION

In this section, we assess commonly used biometric traits against assessment criteria and their accuracy using FAR and FRR. We thoroughly analyze these traits in user authentication applications, considering key performance indicators such as security, convenience, interoperability, scalability, cost, privacy, usability, and robustness. Lastly, we address prevalent cyberattacks targeting these traits and suggest countermeasures to mitigate vulnerabilities.

### A. Evaluation Based on Assessment Criteria

The effectiveness and suitability of various traits in the authentication field are determined by key parameters, known as the Assessment Criteria. This framework, applicable across different authentication methods, includes Universality, Uniqueness, Permanence, Collectability, Performance, Acceptability, and Circumvention.

Universality refers to the prevalence of biometric traits among a certain population. The more universal a biometric trait is, the more potential users it can serve. Uniqueness refers to the distinctiveness of a biometric trait within a population. The uniqueness of a biometric trait is important for accurate identification, as it helps ensure that an individual can be reliably distinguished from others.

Permanence refers to the stability of a biometric trait over time. Permanence is important as it ensures that an individual can be reliably recognized throughout their life, regardless of age or other factors. Collectability refers to the ease of collecting and measuring a biometric trait. Collectability is important as it affects the feasibility and cost of implementing a biometric system . Performance refers to the accuracy and reliability of a biometric system in recognizing an individual based on their biometric trait. The performance of a biometric system is important for ensuring the security and privacy of users, as well as the effectiveness of the system in practical applications. Acceptability refers to the level of public acceptance and willingness to use a biometric system. It takes into account factors such as privacy concerns, cultural beliefs, and the perceived benefits of the system . Circumvention refers to the ease of bypassing or tricking a biometric system. Circumvention is important as it affects the security of a biometric system and its ability to accurately recognize individuals.

In Table III, we conduct a comprehensive evaluation of commonly used traits in authentication applications against these criteria based on related articles such as [192], [193]. This analysis facilitates the selection of the most appropriate authentication method, tailored to the specific requirements and challenges of each application scenario.

### B. Evaluation Based on FAR and FRR

Evaluating biometric authentication systems requires assessing the False Acceptance Rate (FAR) and False Rejection Rate (FRR) metrics, which indicate the probability of false identity acceptance and true identity rejection, respectively. These metrics help determine a system's effectiveness, with lower FAR and FRR values signifying better performance [194].

However, a trade-off exists between these metrics, necessitating a balance for optimal security and user convenience. Factors influencing FAR and FRR include biometric modality, data quality, matching algorithms, and decision-making thresholds, all requiring careful consideration during system design and implementation [195].

According to our extensive review, the following papers [196]–[239] are the ones we found discussing FAR and FRR metrics for the most commonly used biometric traits in user authentication applications. We present the results in Table IV, showing the range for each trait from lowest to highest. A low FAR indicates security, while a low FRR suggests user-friendliness. Conversely, high FAR and FRR values denote reduced security and user-friendliness, respectively. Limited data availability for certain biometric traits (i.e., heartbeat, veins, EAR) results in "NA" entries in the table.

TABLE II: Suitability of Biometric Traits for Different Authentication Applications

| Application | Fingerprint | Face | Iris | Heartbeat | Veins | Ear | Hand Geometry | Retina | Signature | Gait | Voice | Keystroke & Mouse |
|---|---|---|---|---|---|---|---|---|---|---|---|---|
| Static Auth. | ✓ | ✓ | ✓ | ✗ | ✓ | ✓ | ✓ | ✓ | ✓ | ✗ | ✗ | ✗ |
| Remote Auth. | ✗ | ✓ | ✗ | ✓ | ✗ | ✗ | ✗ | ✗ | ✓ | ✗ | ✓ | ✓ |
| Dynamic Auth. | ✗ | ✗ | ✗ | ✓ | ✗ | ✗ | ✗ | ✗ | ✗ | ✗ | ✓ | ✓ |
| Real-time Auth. | ✓ | ✓ | ✗ | ✗ | ✓ | ✓ | ✓ | ✓ | ✓ | ✗ | ✓ | ✗ |
| Continuous Auth. | ✗ | ✗ | ✗ | ✓ | ✗ | ✗ | ✗ | ✗ | ✗ | ✓ | ✓ | ✓ |
| Contact-Based Auth. | ✓ | ✗ | ✗ | ✗ | ✗ | ✗ | ✓ | ✗ | ✓ | ✗ | ✗ | ✗ |
| Contactless-Based Auth. | ✗ | ✓ | ✓ | ✓ | ✓ | ✓ | ✗ | ✓ | ✗ | ✓ | ✗ | ✓ |

TABLE III: Comprehensive Evaluation of Authentication Traits Based on Assessment Criteria

| Biometric traits | Universality | Uniqueness | Permanence | Collectability | Performance | Acceptability | Circumvention |
|---|---|---|---|---|---|---|---|
| Fingerprint | Moderate | High | High | Moderate | High | Moderate | High |
| Face | High | Low | Moderate | High | Low | High | Moderate |
| Iris | High | High | High | Moderate | High | Moderate | Low |
| Heartbeat | High | High | High | Moderate | High | Moderate | Low |
| Veins | Moderate | Moderate | Moderate | Moderate | Moderate | Moderate | Low |
| Ear | High | High | Moderate | Moderate | Moderate | Moderate | Moderate |
| Hand geometry | Moderate | Moderate | Moderate | High | Moderate | Moderate | Moderate |
| Retina | High | High | Moderate | Low | High | Moderate | Low |
| Signature | Low | Low | Moderate | High | Moderate | Moderate | High |
| Gait | Moderate | Low | Low | High | Low | Moderate | Moderate |
| Voice | Moderate | Low | Low | Moderate | Low | High | High |
| Keystroke & Mouse | Low | Low | Low | Moderate | Low | Moderate | Moderate |

TABLE IV: FAR and FRR Comparison Across Biometric Traits

| Biometric Method | Study | FAR | FRR |
|---|---|---|---|
| Fingerprint | [196]–[200] | 0.1% - 0.01% | 0.1% - 1.0% |
| Face | [201]–[205] | 0.1% - 1.0% | 1.0% - 5.0% |
| Iris | [206]–[209] | 0.0001% - 0.01% | 0.1% - 0.5% |
| Heartbeat | NA | NA | NA |
| Veins | NA | NA | NA |
| Ear | NA | NA | NA |
| Hand Geometry | [210]–[214] | 0.1% - 1.0% | 1.0% - 5.0% |
| Retina | [215]–[219] | 0.01% - 0.0001% | 0.1% - 0.5% |
| Signature | [220]–[224] | 1.0% - 5.0% | 1.0% - 5.0% |
| Gait | [225]–[229] | 1.0% - 5.0% | 1.0% - 5.0% |
| Voice | [230]–[234] | 1.0% - 5.0% | 1.0% - 5.0% |
| Keystroke & Mouse | [235]–[239] | 0.1% - 1.0% | 1.0% - 5.0% |

*C. Evaluation Based on Key Performance Indicators*

To evaluate the effectiveness of biometric authentication, it is important to consider various key performance indicators, i.e., security, convenience, interoperability, scalability, cost, privacy, usability, and robustness.

The security of each method is evaluated based on its ability to accurately and reliably identify individuals, while the convenience is evaluated based on its ease of use for the users [240]. The interoperability is evaluated based on the ability of the method to work with different systems and devices, and the scalability is evaluated based on its ability to accommodate large numbers of users.

The cost is evaluated based on the monetary expenses involved in deploying and using the method, while privacy is evaluated based on the protection of personal information. The usability is evaluated based on the ease of use for the users, and the robustness is evaluated based on the method's ability to withstand various attacks and errors [241].

In Table V, we present a table that provides a high-level evaluation of these traits for the commonly used biometric traits in user authentication applications against the selected eight key performance indicators based on three major sources:

- Literature review: We collect the available results from related articles such as [242], [243].
- Online searching: We further collect relevant information from the research databases and industry reports.
- Our own observations: Based on the collected information, we discuss the final results based on our own observations.

*D. Common Cyber Attacks*

This subsection provides an overview of cyber attacks commonly targeting the commonly used biometric traits in user authentication systems. Specifically, we focus on attacks that solely target these traits, rather than those targeting the entire system.

*1) Deepfake Attack:* Deepfake cyber attacks utilize AI and ML to create convincing fake biometric samples for unauthorized access to sensitive data, posing significant security risks. Countermeasures include liveness detection, multi-factor authentication, and deepfake detection technology [244]–[246]. A DCNN-based visual speaker authentication scheme, comprising the FFE-Net and RC-Net, is proposed in [247] and demonstrates robust performance against DeepFake attacks on GRID and MOBIO datasets.

[248] introduces a ResNet50 and Spatial Pyramidal Pooling-based deep fake image detection model, achieving 94% accuracy in real-time data for enhanced facial biometric authentication security. Furthermore, [249] proposes a secure, biometric-based digital ID system for IoT environments using a CNN-based feature extraction method to generate a bio-key from facial features and username, ensuring secure access to smart city facilities.

*2) Template Attack:* The Template cyber attack poses a significant risk by stealing or altering biometric templates, enabling impersonation and unauthorized access to sensitive data. To mitigate this, encryption and access controls can be implemented to limit template access to authorized users only

TABLE V: Comparative Analysis of Biometric Traits Across Key Performance Indicators

| Biometric Traits | Security | Convenience | Interoperability | Scalability | Cost | Privacy | Usability | Robustness |
|---|---|---|---|---|---|---|---|---|
| Fingerprint | High | High | Moderate | High | Moderate | Moderate | High | High |
| Face | Moderate | High | High | High | Low | Low | High | High |
| Iris | High | Moderate | High | High | High | High | High | High |
| Heartbeat | High | Low | Low | Low | High | High | Low | Moderate |
| Veins | High | Low | Low | Low | High | High | Low | High |
| Ear | Moderate | Moderate | Low | Low | High | High | Low | High |
| Hand Geometry | Moderate | High | High | High | Moderate | High | High | High |
| Retina | High | Low | High | Low | High | High | Low | High |
| Signature | Low | High | High | High | Low | High | High | High |
| Gait | Low | High | Low | Low | High | High | High | High |
| Voice | Low | High | High | High | Low | High | High | High |
| Keystroke & Mouse | Low | High | High | High | Low | High | High | High |

[250]–[252]. Techniques such as fuzzy vaults and cancelable biometrics provide additional protection, with fuzzy vaults adding noise to templates and cancelable biometrics employing mathematical transformations to render stolen templates useless [253].

However, cancelable biometrics may be vulnerable to similarity-based attacks, as demonstrated by Particle Swarm Optimization's (PSO) limited applicability in attacking these schemes [254]. Addressing security concerns in brain biometrics using electroencephalography (EEG) data, [255] proposes the first cancellable EEG template design based on a deep learning model and non-invertible transform, effectively protecting raw EEG data and resisting multiple attacks.

*3) Template Update Attack:* The Template Update cyber attack involves an attacker replacing stored biometric templates with their own data to impersonate legitimate users. Countermeasures include liveness detection, regular updates and secure storage of templates, and monitoring unusual authentication patterns [256]–[258]. Biometric backdoors, a template poisoning attack, are studied in [259], revealing adversaries can perform such attacks with physical limitations and zero knowledge of training data, necessitating a poisoning detection technique.

[260] proposes a linear offset-based poisoning attack method, "LOPA," against online self-update fingerprint systems, effectively reducing authentication performance by 42.86%. Lastly, [261] employs blockchain technology for fault-tolerant access and security of deep learning models and biometric templates, alerting the entire system to potential tampering.

*4) Side-channel Attack:* Side-channel cyber attacks threaten biometric systems by exploiting their physical properties to bypass authentication. Implementing physical security measures, such as tamper-resistant hardware and encrypted data storage and transmission, is crucial to protect against these attacks. Intrusion detection systems can monitor for unusual activity, while secure communication channels ensure data access by authorized parties only. With biometric authentication becoming increasingly prevalent, addressing side-channel vulnerabilities is imperative [262]–[264].

The paper, [265], presents models for leakage and attacks against Bozorth3 and custom matching algorithms. The 'Unified Side-Channel Attack Model' (USCA-M) [266] categorizes vulnerabilities and exploit techniques, aiding in the development of defense strategies. Moreover, [267] addresses gait-based key generation vulnerability by proposing a binary classifier combining deep Convolutional Neural Network models and Generative Adversarial Networks, achieving a 97.82% accuracy in preventing video-based gait measurements from generating security keys.

*5) Presentation Attack:* The Presentation cyber attack deceives biometric systems by presenting fake samples, potentially breaching sensitive data, such as bank accounts or medical records. Effective countermeasures include liveness detection (e.g., eye tracking, voice recognition, or facial movement detection) and multi-factor authentication, combining biometrics with passwords or tokens [268]–[270].

A novel fingerprint presentation attack detection (PAD) scheme utilizes a capture device acquiring short wave infrared images and deep learning-based analysis of handcrafted features, achieving a low 1.35% detection error rate on a database of over 4700 samples [271].

Another paper discusses face PAD, biometric system vulnerabilities, attack methods, detection techniques, and public datasets used in research [272]. A new PAD approach employs a light field camera, capturing multiple depth images simultaneously, and demonstrates outstanding performance on a database of simulated face artefacts [273]. Despite numerous PAD techniques, sophisticated presentation attacks remain a growing concern for face recognition systems.

*6) Replay Attack:* Replay cyber attacks deceive authentication systems by capturing and later replaying valid biometric samples, granting unauthorized access to sensitive information and causing significant damage. To counteract these attacks, challenge-response protocols ensure fresh biometric samples, and multi-factor authentication requires additional authentication factors, such as passwords, tokens, or smart cards. Various approaches have been proposed to mitigate such attacks.

In [274], the authors introduced Pistis, an authentication protocol integrating gait biometrics and liveness detection for wearable gait biometric systems. Also, the paper, [275], presented a protocol for dynamics exchange in keystroke dynamics using real and fake information snippets, substantially reducing replay attack possibilities while maintaining accurate user verification.

Lastly, [276] proposed a non-deterministic approach for iris recognition, employing randomly selected subsets of robust iris regions for authentication, effectively mitigating replay attacks and withstanding hill climbing attacks.

*7) Insider Attack:* The Insider cyber attack allows authorized individuals to exploit biometric data for malicious purposes, bypassing authentication measures and posing a significant threat to system security. Detection and prevention are challenging due to the attacker's seemingly authorized status, enabling undetected access to sensitive data or systems. Mitigation strategies include strict access controls, auditing procedures, and anomaly detection techniques, which help identify abnormal system usage patterns and prevent unauthorized access [277]–[279].

In [280], the EAPIOD authentication protocol is proposed for IoD protection against insider attacks, with security analyses confirming its immunity and superior performance compared to existing schemes. The paper, [281], introduces a trusted platform module-based authentication framework for Hadoop, offering robust security against insider threats, and its effectiveness is supported by experiments and formal proof. Finally, [282] presents a reputation score-based model utilizing blockchain technology for detecting post-authentication attacks in IoT infrastructure, providing a secure and lightweight solution for smart metering systems.

*8) Social Engineering Attack:* Social Engineering is a cyber attack that tricks users into divulging biometric data or compromising the biometric system through non-technical means such as phishing, pretexting, or baiting. These deceptive methods can bypass authentication or grant unauthorized access to sensitive information. To counter these attacks, it is crucial to educate users on social engineering risks, enforce strict access control policies, employ multi-factor authentication, and regularly monitor and audit the biometric system.

Various papers propose methods to detect and prevent phishing attacks [283]–[285]. One study [286] presents a model for detecting and preventing Social Engineering Based Phishing Attacks (SEBPA) on Facebook, validated through four realistic scenarios, and predicts threatening situations using color-coded warnings. Another paper [287] proposes a neural network-based phishing prevention algorithm (PPA) implemented with Ryu, an open-source SDN controller, successfully detecting phished versions of Facebook, Yahoo, and Hotmail login pages.

Lastly, an in-depth phishing model [288] considers attack stages, attacker types, threats, targets, channels, and tactics, utilizing an AI-based LSTM detection method that demonstrates satisfactory performance and accuracy. Table VI summarizes the findings from this subsection.

## V. Discussion

This section aims to provide a comprehensive overview of the findings presented in this paper, analyzing the key factors influencing the accuracy and effectiveness of biometric systems, as well as the advantages and disadvantages of commonly used biometric traits in user authentication applications. In this section, we will also address the potential future research directions.

### A. Influential Factors

Several factors can affect the accuracy and effectiveness of biometric systems. Among the common influential factors are age, disease, illuminations, injury, noise, fatigue, and apparel changes. To determine the susceptibility of the commonly used biometric traits to these factors, this section will provide a comparison.

*1) Fingerprint Recognition:* There are several factors that can affect fingerprint its performance, such as age, disease, and injury. As people age, their skin loses elasticity and the ridges and valleys of their fingerprints become less distinct, which can result in a decline in accuracy and an increase in false negatives [289].

Certain diseases, such as eczema and psoriasis, can change the skin texture and reduce the accuracy of the system. Injuries, such as cuts and burns on the fingers, can also alter the texture and ridges, causing mismatches between the stored image and the newly captured image. For instance, an elderly person with less defined fingerprints or a person with eczema on their fingers may have difficulty accessing their device or secure facilities using fingerprint authentication.

*2) Face Recognition:* Face recognition, a popular user authentication method, is impacted by various factors such as age, disease, illumination, injury, and apparel changes. Aging can cause facial features to become less defined, while wrinkles and sagging skin further affect recognition accuracy. Diseases like acne, scarring, or facial surgery also challenge the system's ability to identify individuals.

Lighting conditions play a crucial role; poor lighting can result in blurry images, making accurate identification difficult [290]. Injuries like cuts, burns, or bruises on the face and changes in apparel, such as hats or glasses, can similarly alter one's appearance, complicating the face recognition process.

*3) Iris Recognition:* Iris is not immune to Several factors can impact the accuracy of iris recognition. Disease, such as cataracts or glaucoma, can change the appearance of the iris and affect its recognition. Scarring in the iris can also reduce recognition accuracy. Illumination plays a critical role in iris recognition. Poor lighting conditions can affect image quality, resulting in incorrect or failed authentication [291].

Apparel changes, such as glasses or headwear, can also impact the accuracy of iris recognition. These factors must be taken into account to ensure reliable and effective biometric authentication.

*4) Heartbeat Recognition:* Heartbeat recognition accuracy can be affected by various factors such as age, injury, disease, and fatigue. Age is significant since the heart changes over time; heart rate tends to slow down, and the rhythm can become irregular, making it challenging for the authentication system to accurately recognize an older person's heartbeat.

Injuries, such as heart attacks, and diseases like heart arrhythmias and heart failure can also affect recognition accuracy by changing the heartbeat pattern [292]. Fatigue is another factor that can cause changes in the heartbeat pattern, leading to recognition errors. For instance, a person's heartbeat may become slower when tired, making it difficult for the authen-tication system to accurately identify their identity.

TABLE VI: Overview of Cybersecurity Threats to Biometric Traits

| Attack Type | Description | Impact | Countermeasures |
|---|---|---|---|
| Deepfake | Using AI and ML to create realistic fake biometric samples, such as voice, face, or speech, to bypass the authentication system | Bypass authentication | Use anti-spoofing techniques such as liveness detection or employ multi-factor authentication that includes biometrics and other factors such as passwords or tokens. Employ deepfake detection technology that can identify fake biometric samples. |
| Template | Stealing biometric templates stored in the system's database to impersonate a user, or modifying stored templates to match an attacker's biometric pattern | legitimate users Impersonation | Use encryption and access controls to protect stored templates, or employ template protection techniques such as fuzzy vaults or cancelable biometrics |
| Template Update | Attacker updates the biometric template stored in the system with their own biometric data | legitimate users Impersonation | Use of liveness detection to ensure that the biometric being presented is from a live user and not a stored template. Regular updates and secure storage of biometric templates. Monitoring for unusual authentication patterns |
| Side-channel | Attacker analyzes the physical properties of the biometric sensor or system to deduce the biometric template or user's biometric data | Bypass authentication | Implementing physical security measures to protect the biometric sensor and system, such as tamper-resistant hardware or encryption of biometric data in transit and at rest. Using intrusion detection systems to monitor for unusual physical activity. Using secure communication channels between the biometric sensor and the system. |
| Presentation | Presenting a fake biometric sample to the system to deceive it, using techniques such as 3D printing or artificial skin | Bypass authentication | Use liveness detection to ensure that the biometric sample is from a live person, or employ multi-factor authentication that combines biometrics with other factors such as passwords or tokens |
| Replay | Capturing a valid biometric sample during an authentication session and replaying it to the system later | Bypass authentication | Use challenge-response protocols to ensure that the biometric sample is fresh, or employ multi-factor authentication |
| Insider | Authorized individuals using their own or another user's biometric data to bypass authentication for malicious purposes | Bypass authentication | Implement strict access controls and auditing procedures, or employ anomaly detection to detect abnormal patterns of system usage |
| Social Engineering | Tricking users into divulging their biometric data or compromising the biometric system through non-technical means, such as phishing, pretexting, or baiting | Bypass authentication or gain unauthorized access | Educate users on the risks of social engineering attacks and how to identify and avoid them. Implement strict access control policies to limit access to sensitive biometric data. Employ multi-factor authentication that includes biometrics and other factors such as passwords or tokens. Regularly monitor and audit the biometric system for suspicious activity. |

*5) Veins Recognition:* Vein recognition technology, like any biometric system, faces challenges that can impact its accuracy. Injury is a major challenge that can alter the pattern of veins in a person's finger, resulting in authentication failure [293]. Even if a broken finger has healed, the system may not be able to recognize the person due to the changed veins. Moreover, certain diseases such as rheumatoid arthritis or leukemia can also cause changes in the pattern of veins, leading to authentication failure.

*6) Ear Recognition:* Ear recognition accuracy and reliability in real-world scenarios can be affected by various factors, resulting in false negative or positive identifications, which can lead to security breaches or denied access for legitimate users. Age can impact ear recognition, as the ear shape and structure of individuals can change over time, making it more difficult for recognition systems to identify them accurately [100].

Medical conditions such as skin diseases or injuries like cuts or burns can alter the ear's appearance and structure, leading to inaccurate recognition. Poor lighting conditions can also hinder recognition accuracy, as can changes in apparel such as hats, hoods, or earmuffs, which can obstruct the ear's view.

*7) Hand Geometry :* This method is limited by changes in an individual's hand due to aging, disease, injury, and apparel changes [188]. Aging affects all living organisms, shrinking hand size, and making bones less dense, altering unique characteristics. False rejection or acceptance can occur, and even temporary injury like a cast can affect accuracy. Similarly, disease and injury like arthritis, carpal tunnel syndrome, or broken bones can alter hand shape and size, leading to recognition difficulties. Lastly, apparel changes, such as gloves or hand coverings, can obstruct the view and lead to false acceptance or rejection, causing potential security breaches.

*8) Retina Recognition:* Retina recognition effectiveness is affected by age, disease, and injury. The retina changes over time as a person ages, becoming more transparent and making patterns less distinct, thus challenging recognition accuracy [116]. Age-related diseases such as macular degeneration can worsen the deterioration of the retina, further impacting the technology's performance.

Diseases like diabetic retinopathy, glaucoma, and cataracts can damage the retina, alter its patterns, and reduce its light sensitivity, making it even more challenging for recognition. Injury is another factor that can change the retina's structure and appearance. Trauma can cause detachment and shift in position of the retina, while inflammation can alter its appearance and decrease its recognition accuracy.

*9) Signature Recognition:* Signature recognition accuracy can be affected by various factors such as age, disease, injury, and fatigue [294]. Age can reduce fine motor control and dexterity, causing variations in signature quality, particularly with size and spacing of letters. Disease, such as Parkinson's or neurological disorders, can also lead to changes in signature due to difficulty with fine motor control.

Injury to the hand, wrist, or arm can affect signature writing ability, while fatigue and stress can lead to variations in signature precision. These factors can cause discrepancies between a person's recorded signature and what the recognition system identifies, making accurate identification difficult.

*10) Gait Recognition:* Gait recognition is a non-intrusive method that has gained popularity due to its unique features. However, it can be affected by various factors such as age, disease, illuminations, injury, fatigue, and apparel changes [295]. For instance, age can cause changes in gait patterns, leading to a reduction in accuracy. Similarly, diseases like Parkinson's or multiple sclerosis can impact gait recognition due to tremors or stiffness.

Illuminations, such as low light conditions, and changes in apparel type can also affect gait recognition accuracy. Fatigue and injury can alter gait patterns, causing incorrect authentication. Therefore, it's essential to consider the effect of apparel changes when developing gait recognition systems.

*11) Voice Recognition:* Voice recognition, also known as speech recognition, is a biometric authentication method that is not immune to factors that can affect its accuracy and reliability. Disease, such as a cold or throat infection, can temporarily change a person's voice, leading to false rejections [296]. Injury to the throat, such as laryngeal fractures, and to the mouth, such as jaw fractures, can also impact a person's speech and cause false rejections.

Background noise can interfere with the sound of a person's voice and cause the recognition system to misinterpret speech patterns, leading to false rejections or acceptances. Lastly, fatigue can change a person's voice, making it difficult for the recognition system to accurately identify them and causing false rejections, which can make it challenging for individuals to access secure information or unlock their devices.

*12) Keystroke and Mouse Movement:* Keystroke and mouse movement are impacted by various factors such as age, disease, injury, and fatigue. Age is one of the most significant factors that affect user authentication, as physical and cognitive declines can impact typing speed, accuracy, and mouse movements. Parkinson's and multiple sclerosis can cause tremors and muscle weakness, making it challenging for users to enter the correct authentication information [162].

Injury, particularly to the hands, can also have a significant impact on keystroke and mouse movement, while fatigue can affect typing speed and accuracy, leading to a higher rate of authentication failure. These factors can reduce the usability of these methods, particularly for older users, and may require alternative authentication methods that don't require physical input.

In Table VII, we summarize the outcome of this subsection compares various biometric traits and their vulnerability to different influential factors. The factors considered are age, disease, illuminations, injury, noise, fatigue, and apparel changes. The table uses the symbols ∨ (affects) and ✗ (does not affect) to represent the influence of each factor on each biometric trait.

### B. Advantages and Disadvantages

In this subsection, we delve into the specific advantages and disadvantages of the biometric traits previously discussed. For a detailed summary of these attributes for physiological biometric traits, refer to Table VIII. The advantages and disadvantages of behavioral biometric traits are summarized in Table IX.

### C. Qualitative Comparison

Answering our research questions sets our paper apart from recently published surveys [297]–[302] in the following ways:

1) Unlike [297]–[302], our paper explores both physiological and behavioral biometric traits commonly used in user authentication applications, making our coverage more extensive compared to other surveys.
2) Our survey paper differs from previous studies [298], [299], [301], [302] in that they focus solely on the development of biometric user authentication on mobile phones, whereas our paper covers biometric user authentication in general.

Table X compares our survey against recent published surveys [297]–[302] and lists the twelve most commonly used biometric traits in user authentication applications. The table indicates whether each survey covers the biometric traits or not, represented by ∨ (covered) or ✗ (not covered).

Moreover, in our survey, we present a distinctive and comprehensive analysis of biometric traits in user authentication applications, setting it apart from recent studies. Key differentiators of our work include:

TABLE VII: Comparative Impact Analysis of Influential Factors on Different Biometric Traits

| Biometric trait | Age | Disease | illuminations | Injury | Noise | Fatigue | Apparel Changes |
|---|---|---|---|---|---|---|---|
| Fingerprint | ✓ | ✓ | ✗ | ✓ | ✗ | ✗ | ✗ |
| Face | ✓ | ✓ | ✓ | ✓ | ✗ | ✗ | ✓ |
| Iris | ✗ | ✓ | ✓ | ✓ | ✗ | ✗ | ✓ |
| Heartbeat | ✓ | ✓ | ✗ | ✓ | ✗ | ✓ | ✗ |
| Veins | ✗ | ✓ | ✗ | ✓ | ✗ | ✗ | ✗ |
| Ear | ✓ | ✓ | ✓ | ✓ | ✗ | ✗ | ✗ |
| Hand geometry | ✓ | ✓ | ✗ | ✓ | ✗ | ✗ | ✓ |
| Retina | ✓ | ✓ | ✗ | ✓ | ✗ | ✗ | ✗ |
| Signature | ✓ | ✓ | ✗ | ✓ | ✗ | ✓ | ✗ |
| Gait | ✓ | ✓ | ✓ | ✓ | ✗ | ✓ | ✓ |
| Voice | ✗ | ✓ | ✗ | ✓ | ✓ | ✓ | ✗ |
| Keystroke & Mouse | ✓ | ✓ | ✗ | ✓ | ✗ | ✓ | ✗ |

TABLE VIII: Summary of Advantages and Disadvantages: Physiological Traits

| Biometric trait | Advantages | Disadvantages |
|---|---|---|
| Fingerprint | - **Universality:** Almost every individual has a unique set of fingerprints, which makes it an excellent biometric trait for identification and authentication purposes.<br>- **Accuracy:** Fingerprint recognition is highly accurate and reliable, and it is difficult to fake or replicate a person's fingerprints.<br>- **Convenience:** Fingerprint recognition does not require the user to remember passwords or carry authentication tokens. | - **Hygiene concerns:** Fingerprint sensors can be a potential source of germs and bacteria, especially in public environments, which may pose a risk to the user's health.<br>- **Limited universality:** In rare cases, some people may not have identifiable fingerprints due to certain medical conditions or occupational hazards, which makes it unsuitable as a universal biometric trait. |
| Face | - **Non-intrusive:** Face recognition is a non-invasive method of authentication that does not require the user to perform any special actions or procedures.<br>- **Universality:** Almost every individual has a face, which makes it a universally available biometric trait for authentication purposes.<br>- **Convenience:** Face recognition is a fast and convenient method of authentication that can be completed in a few seconds, which makes it convenient for both the user and the system. | - **Twins issue:** Identical twins share similar facial features and physical characteristics, which can make it difficult for the system to differentiate between them.<br>- **Facial expression:** The accuracy of face recognition systems can be affected by a range of facial expressions, such as frowns, and smiles.<br>- **Limited Usage:** Facial recognition may not be suitable for all authentication scenarios, as it may not be reliable in certain situations such as when users are wearing masks. |
| Iris | - **Fast:** The process of capturing an image of the iris and matching it with a pre-stored template can take just a few seconds.<br>- **Difficult to duplicate or forge:** Iris is an internal organ that is protected by the cornea and is not readily visible or accessible.<br>- **Long-term stability:** The iris is a stable biometric trait that does not change significantly over time, making it suitable for long-term authentication purposes. | - **Equipment limitations:** Iris recognition requires specialized equipment, such as a high-resolution camera and a dedicated software system, which can be costly and may require trained personnel to operate.<br>- **Close proximity:** Iris recognition systems require the user to position their eyes within a certain distance and orientation from the scanner to capture a high-quality image of the iris. |
| Heartbeat | - **Unique:** The heartbeat pattern is unique to an individual, making it a good biometric trait for identification and authentication purposes.<br>- **Difficult to fake:** It is difficult to replicate a person's heartbeat pattern, making it a strong deterrent against fraud and impersonation. | - **Technical challenges:** Measuring heartbeat requires specialized equipment and software, which may not be available or affordable.<br>- **Environmental factors:** Heartbeat can be affected by environmental factors such as physical activity, or illness, which may lead to false negatives or false positives in authentication. |
| Veins | - **Difficult to forge:** Vein patterns are difficult to replicate, making it impossible for an unauthorized person to access the system.<br>- **Hygienic:** Vein recognition does not require physical contact, reducing the risk of spreading diseases or infections.<br>- **Longevity:** Vein patterns do not change significantly over time, ensuring consistent and reliable authentication over an extended period. | - **Expensive:** Vein recognition technology can be costly to implement, making it less accessible to smaller organizations.<br>- **Limited availability:** The technology is still in the early stages of development and not widely available in many regions.<br>- **User discomfort:** Some individuals may experience discomfort or anxiety when using vein recognition technology, which could result in poor user adoption. |
| Ear | - **Non-intrusive:** Ear recognition is a non-intrusive biometric technology that does not require physical contact with the user, making it more comfortable and hygienic than other biometric technologies that require users to touch sensors.<br>- **Difficulty to fake:** The shape and features of ears are difficult to fake, which makes ear recognition a more secure authentication method. | - **Equipment requirements:** Ear recognition requires specialized equipment, such as high-resolution cameras, to capture and analyze ear images, which may increase the cost of implementation.<br>- **Limited adoption:** Ear recognition is not as widely adopted as other biometric technologies, such as fingerprint or facial recognition, which may limit its usefulness in some settings. |
| Hand geometry | - **Ease of use:** An individual being authenticated simply places their hand on a scanner, and the system automatically captures the required measurements.<br>- **Suitable for Harsh Environments:** Hand geometry biometrics can function effectively in harsh environments, such as dusty or dirty workplaces, making it a practical choice for industrial settings. | - **Limited universality:** Hand geometry may not be a universal biometric trait, as some individuals may have physical conditions or injuries that affect the measurements of their hand, making it difficult to authenticate them using this trait.<br>- **Limited Scalability:** Hand geometry biometrics can only authenticate one user at a time, making it unsuitable for organizations with large numbers of users. |
| Retina | - **Unique:** Each person's retina patterns are unique, and cannot be easily duplicated or forged. This makes retina recognition a highly secure method of user authentication.<br>- **Stable:** Retina patterns remain relatively stable throughout a person's life, so a single registration can serve for many years. This can make it a cost-effective method of user authentication. | - **Technical challenges:** Retina recognition requires sophisticated hardware and software, as well as highly trained personnel, to properly implement and maintain.<br>- **Close proximity:** Retina recognition requires a user to be in very close proximity to the scanning device, as it must be captured with a high level of precision in order to be accurately identified. |

TABLE IX: Summary of Advantages and Disadvantages: Behavioral Traits

| Biometric trait | Advantages | Disadvantages |
|---|---|---|
| Signature | - **Easy to use:** Signatures are easy to use and require no special equipment, making it a convenient biometric trait.<br>- **Low barrier to entry:** Most people are already familiar with signing their name, making signature authentication an accessible method for users of all ages and technical abilities. | - **Forgery:** Signatures can be forged, making them less secure than other biometric traits.<br>- **Lack of standardization:** There are no standardized methods for signature authentication, making it difficult to compare results across different systems. |
| Gait | - **Non-Intrusive:** Gait analysis can be performed remotely, without requiring any specialized hardware or sensors.<br>- **Difficult to mimic:** It is difficult to mimic someone else's gait pattern.<br>- **Passive Authentication:** Gait analysis can be performed without requiring any active participation from the user. | - **Environmental Factors:** Gait analysis is sensitive to environmental factors like lighting, terrain, and footwear.<br>- **Limited Performance in Crowded Areas:** In crowded areas, it may be challenging to capture an individual's gait pattern accurately due to the presence of other people and obstacles. |
| Voice | - **Easy to Use:** Voice biometric authentication is easy to use, requiring only a spoken phrase or a few words to authenticate the user, making it accessible to a wide range of users.<br>- **Non-Invasive:** Voice biometric authentication is non-invasive and does not require physical contact, making it a convenient and user-friendly option for authentication. | - **Impersonation:** Although voice is unique, an individual may still be able to impersonate another person's voice, especially if they have a similar voice pattern.<br>- **Environmental Factors:** The voice biometric system may fail to recognize an individual's voice due to environmental factors, such as background noise or poor voice quality. |
| Keystroke & Mouse | - **User-friendly:** Keystroke and mouse biometrics are easy to use and convenient, as users do not have to remember complex passwords or carry physical tokens with them.<br>- **Non-intrusive:** Keystroke and mouse biometrics do not require any physical contact, making it a non-intrusive and easy-to-use method of authentication. | - **Limited scope:** Keystroke and mouse biometrics can only be used for user authentication in online systems, limiting its scope of use to internet-based applications and services.<br>- **External factors affect accuracy:** External factors such as hand injuries, fatigue, and changes in typing habits can affect its accuracy, making it less reliable. |

TABLE X: Comparative Coverage Analysis of Biometric Traits in Our Survey and Recent Published Surveys

| Study | Fingerprint | Face | Iris | Heartbeat | Veins | Ear | Hand Geometry | Retina | Signture | Gait | Voice | Keystroke & Mouse |
|---|---|---|---|---|---|---|---|---|---|---|---|---|
| [297] | ✓ | ✓ | ✗ | ✗ | ✗ | ✗ | ✓ | ✓ | ✗ | ✗ | ✓ | ✗ |
| [298] | ✓ | ✓ | ✓ | ✗ | ✗ | ✗ | ✗ | ✗ | ✗ | ✓ | ✓ | ✓ |
| [299] | ✗ | ✗ | ✗ | ✗ | ✗ | ✗ | ✗ | ✗ | ✗ | ✓ | ✗ | ✓ |
| [300] | ✓ | ✓ | ✓ | ✗ | ✗ | ✗ | ✓ | ✗ | ✓ | ✗ | ✓ | ✓ |
| [301] | ✗ | ✗ | ✗ | ✗ | ✗ | ✗ | ✗ | ✗ | ✓ | ✓ | ✓ | ✓ |
| [302] | ✓ | ✓ | ✓ | ✗ | ✓ | ✗ | ✓ | ✓ | ✓ | ✓ | ✓ | ✓ |
| Ours | ✓ | ✓ | ✓ | ✓ | ✓ | ✓ | ✓ | ✓ | ✓ | ✓ | ✓ | ✓ |

1) We delve into the suitability of various biometric traits for user authentication, a dimension not fully explored in studies such as [297]–[302].
2) Our survey uniquely evaluates biometric traits against criteria like Universality, Uniqueness, Permanence, Collectability, Performance, Acceptability, and Circumvention, a perspective not thoroughly covered in [298], [299], [301].
3) We incorporate an accuracy assessment using FAR and FRR metrics, aspects not addressed in [297], [300].
4) Our survey offers an in-depth analysis of performance indicators like security, convenience, interoperability, scalability, cost, privacy, usability, and robustness. This is a comprehensive approach not seen in [297]–[302].
5) We specifically address cyber attacks targeting biometric traits in authentication systems and propose countermeasures, a focus area generally overlooked in [297]–[302].
6) Unlike other surveys, we examine how various factors like age, disease, illumination, injury, noise, fatigue, and apparel changes affect biometric traits, offering insights not found in the aforementioned studies.
   We provide a detailed exploration of the specific pros and cons of biometric traits, an area not sufficiently addressed in [297], [298], [301].

Table XI effectively showcases our survey's unique contributions to biometric authentication research through a clear, comparative framework. Utilizing a checkmark (✓) or cross (✗) system, it highlights the presence or absence of key concepts in comparison to other studies. This visual representation underlines the comprehensive depth and breadth of our work, emphasizing its significant advancements beyond the scope of existing literature. Our distinct approach not only sets our survey apart but also establishes it as an essential resource in the field, marking a substantial leap in biometric authentication study.

### D. Future Research Directions

Although the biometric user authentication application field has made significant progress, there are still many areas that require further research and development.

*1) Artificial Intelligence and Machine Learning Application:* Artificial Intelligence (AI) and Machine Learning (ML) have the potential to revolutionize the field of biometric user authentication. Researchers are exploring the use of ML and AI algorithms to enhance the accuracy and reliability of biometric authentication systems. For example, deep learning algorithms can be trained to recognize and verify multiple biometric traits simultaneously. Moreover, reinforcement learning techniques can be used to improve the performance of continuous authentication systems by adapting to the user's behavior over time.

Another potential area of research is the development of biometric systems that can adapt to changes in the user's biometric traits over time. For example, as a person ages, their facial features may change, making it more difficult for a facial recognition system to identify them. By using ML algorithms to analyze and learn from the user's biometric data over time, a biometric system can adapt to these changes and maintain a high level of accuracy and reliability.

TABLE XI: Comparative Overview of Evaluation Concepts in Our Survey and Recent Published Surveys

| Study | Application Suitability | Assessment criteria | FAR & FRR | Key Performance Indicators | Common Cyber Attacks | Influential Factors | Advantages and Disadvantages |
|---|---|---|---|---|---|---|---|
| [297] | ✗ | ✓ | ✗ | ✗ | ✗ | ✗ | ✗ |
| [298] | ✗ | ✗ | ✓ | ✗ | ✗ | ✗ | ✗ |
| [299] | ✗ | ✗ | ✓ | ✗ | ✗ | ✗ | ✓ |
| [300] | ✗ | ✗ | ✗ | ✗ | ✗ | ✗ | ✓ |
| [301] | ✗ | ✗ | ✓ | ✗ | ✗ | ✗ | ✗ |
| [302] | ✗ | ✓ | ✓ | ✗ | ✗ | ✗ | ✓ |
| Ours | ✓ | ✓ | ✓ | ✓ | ✓ | ✓ | ✓ |

*2) Adaptive Biometric Systems:* Adaptive biometric systems dynamically adjust their parameters and decision-making processes based on factors such as user behavior, environmental conditions, and contextual information. This can lead to more personalized authentication experiences, as well as increased security and convenience. Future research can explore new algorithms and models for context-aware biometric systems, as well as methods for fusing contextual information with biometric data during the authentication process. Also, studies can investigate the impact of various factors, such as user demographics or device types, on the performance of adaptive biometric systems.

*3) Privacy-Preserving Techniques:* Privacy-preserving techniques and mechanisms protect users' biometric data from unauthorized access and misuse. As privacy concerns grow, it becomes increasingly important to develop innovative methods for secure storage, processing, and transmission of biometric data. Future research can explore novel cryptographic schemes, such as homomorphic encryption or secure multi-party computation, that allow biometric data to be processed securely without revealing sensitive information. Moreover, studies can investigate the application of differential privacy techniques to biometric systems, enabling the analysis of biometric data while maintaining user privacy.

*4) Biometric Authentication in IoT:* The increasing adoption of Internet of Things (IoT) devices necessitates research into biometric authentication solutions suitable for resource-constrained environments. Future work can focus on developing lightweight, energy-efficient, and secure biometric authentication methods for IoT systems, which may involve investigating new biometric traits or optimizing existing methods for IoT contexts. Furthermore, studies can explore the integration of biometric authentication with other IoT security mechanisms, such as device attestation or secure boot processes. Furthermore, research can investigate the challenges and requirements for implementing biometric authentication solutions in various IoT application domains, such as smart homes, healthcare, or industrial control systems.

*5) Standardization, Ethical and Legal Considerations:* Another challenge facing biometric authentication systems is the lack of standardization. There is a need for common standards and protocols to ensure interoperability between different systems and devices. Moreover, standardization can help to address concerns around privacy and security by providing guidelines for the collection, storage, and use of biometric data.

Futhermore, the use of biometric data raises ethical and legal concerns around privacy, security, and human rights. Future research should focus on addressing these concerns by developing guidelines and best practices for the collection, storage, and use of biometric data. Moreover, legal frameworks should be established to regulate the use of biometric data and ensure that the rights of individuals are protected.

## VI. CONCLUSION

This paper has provided a comprehensive survey of biometric user authentication applications, their evaluation, discussion, and future research directions. Through a literature review, online searching, and our own observations, we have answered the research questions posed in the introduction section.

Moreover, we have explored both physiological and behavioral biometric traits commonly used in user authentication applications, their performance in terms of security, convenience, interoperability, scalability, cost, privacy, usability, and robustness, and the key factors influencing the accuracy and effectiveness of biometric systems.

In addition, we have provided a thorough discussion of the advantages and disadvantages of these biometric traits and addressed the crucial aspect of mitigating common cyber attacks that target these traits. Our survey has distinguished itself from previously published surveys in several ways, as outlined in the discussion section.

Finally, biometric authentication is an essential element of security across various domains, and our survey has highlighted the strengths, weaknesses, and future research directions for this critical technology. We believe that our survey can serve as a valuable resource for researchers, practitioners, and decision-makers in the field.